 \documentclass[smallabstract,smallcaptions]{dccpaper}

\usepackage{epsfig}
\usepackage{citesort}
\usepackage{amsmath}
\usepackage{amssymb}
\usepackage{multirow}
\usepackage{color}
\usepackage{url}
\usepackage{subcaption}
\usepackage{algorithm}
\usepackage{algorithmic}

\newlength{\figurewidth}
\newlength{\smallfigurewidth}

\makeatletter
\newcommand*{\rom}[1]{\expandafter\@slowromancap\romannumeral #1@}
\makeatother

\begin{document}

\title
{\large
\textbf{Convex Optimization Based Bit Allocation for Light Field Compression under Weighting and Consistency Constraints}
}

\author{%
Bichuan Guo$^{\ast}$, Yuxing Han$^{\dag}$, Jiangtao Wen$^{\ast}$\\[0.5em]
{\small\begin{minipage}{\linewidth}\begin{center}
\begin{tabular}{ccc}
$^{\ast}$Tsinghua University & \hspace*{0.5in} & $^{\dag}$South China Agricultural University \\
Beijing, 100084, China && Guangzhou, Guangdong, 510642, China\\
\url{jtwen@tsinghua.edu.cn} && \url{yuxinghan@scau.edu.cn}
\end{tabular}
\end{center}\end{minipage}}
}

\maketitle
\thispagestyle{empty}

\begin{abstract}
Compared with conventional image and video, light field images introduce the weight channel,
as well as the visual consistency of rendered view, information that has to be taken into account
when compressing the pseudo-temporal-sequence (PTS) created from light field images. 
In this paper, we propose a novel frame level bit allocation framework for PTS coding.
A joint model that measures weighted distortion and visual consistency, combined with an 
iterative encoding system, yields the optimal bit allocation for each frame by solving a convex optimization problem. 
Experimental results show that the proposed framework is effective in producing desired distortion 
distribution based on weights, and achieves up to 24.7\% BD-rate reduction comparing to the default rate control algorithm.

\end{abstract}

\Section{Introduction}
The light field, introduced in \cite{Gershun}\cite{Adelson},
describes the intensity of light rays passing through each point in space, 
at each possible direction, wavelength and time.
Given a single time instance, 
the light field model can be simplified as a 4D function $\mathcal{L}(u,v,x,y)$,
which can be considered as a collection of perspective images of the $xy$ plane,
observed from various positions on the $uv$ plane \cite{Viola17}.

Recent advancements in light field imaging,
brought to light by the commercial products Lytro and Ratrix,
call for efficient compression algorithms.
\cite{Viola17} evaluated and compared several state-of-the-art light field image compression algorithms,
most of them can be classified as lenslet image compression \cite{Li16}\cite{Conti16} or perspective image compression \cite{Perra16}\cite{Liu16}.
The latter approach arranges perspective images into a pseudo-temporal-sequence (PTS),
which is then coded with the HEVC \cite{sullivan2012overview} video encoder.

Rate control aims at delivering a video stream with the highest possible visual quality,
while keeping the bitrate under constraints.
As PTS coding uses a video encoder,
a rate control mechanism is also required, with 
some important and unique challenges 
not addressed in a general video rate control algorithm design. 
Image captured by a light field camera (e.g. Lytro Illum) may contain a fourth weight channel
in addition to the conventional RGB channels \cite{Vieira2015}, indicating the confidence of each pixel,
and affects the evaluation of coding distortion accordingly.
Moreover, the visual consistency differs from the normal sense,
as the PTS is not displayed in temporal order,
rather, a render algorithm \cite{ng2005light}\cite{levoy1996light} is used to reconstruct views from the PTS.

In this paper, 
we propose a novel frame level bit allocation framework for PTS coding
that takes into account the weighting and consistency factors.
The rate-distortion curves of each perspective frame are estimated
to  deduce the weighted distortion and consistency.
They form a cost function to be minimized, when combined with a total bit cost constraint,
define an optimization problem that can be systematically solved using convex programming,
yielding optimal bit costs of each perspective frame.

The rest of the paper is organized as follows.
Related work is presented in Section \rom{2}.
A quantitative model for weighted distortion and consistency is described in Section \rom{3}.
The convex optimization problem is formulated and solved in Section \rom{4}.
Experimental results are given in Section \rom{5}, and
Section \rom{6} concludes the paper.

\Section{Related Work}

Evaluations in \cite{Viola17} showed that 
perspective image compression is more efficient comparing to lenslet image compression.
This coding approach relies on 
the arrangement of perspective images in the PTS.
Spiral scan order \cite{Viola17} and raster scan order \cite{Perra16} were proposed,
as well as a 2D hierarchical structure \cite{Li17}, 
which achieves higher efficiency with modifications to the HEVC codec.

Recent research \cite{Li2016lambda} proposed using the Lagrange multiplier $\lambda$ 
for bit allocation in HEVC.
\cite{Li17J} proposed a rate control algorithm for its 2D hierarchical 
structure PTS coding based on an analysis of the $R-\lambda$ model.
It achieved significant bitrate savings but did not consider the impact of the 
weighting and consistency factors.

To some extent, rate control in light field image coding is similar to variable bitrate (VBR) coding,
as we do not require the bitrate of the encoded PTS to be constant, but rather, aim to achieve 
the best video quality possible.
Many VBR algorithms employ a two-pass procedure \cite{Yin2004}\cite{Que2006}, 
where encoding related statistics, especially the rate-distortion behavior of each frame, 
is collected during the first pass. These are then
used in the second pass to perform the actual encoding.
However, as the rate-distortion behavior of a frame relies on its reference frames,
either an iterative framework \cite{Hsu1997}, 
which assumes independence within each iteration and seeks convergence by re-encoding,
or a model that features backward dependency \cite{Pang2011} is required.

With frame level rate-distortion behavior, 
one can formulate an optimization problem which minimizes a cost function 
(e.g. distortion, discontinuity, and etc.) while satisfying given constraints,
to yield an optimal solution for bit allocation.
\cite{Sermadevi06} proposed an iterative convex programming framework for 
VBR streaming rate control under multiple channel rate constraints. 
\cite{Pang2013} proposed a convex optimization model with inter-frame dependency 
compatibility to solve the joint bit allocation problem
in H.264 statistical multiplexing.

As its main novelty, this paper proposes an optimization target 
that measures weighted distortion and visual consistency of light field images.
An iterative encoding system is then proposed, 
as well as a two-step strategy that converts the proposed target into a solvable convex optimization problem.

\Section{Weighting and Consistency Models}
\SubSection{Weighted Distortion}
Each pixel in the light field image is associated with a 4-tuple coordinate $(u, v, x, y)$,
as well as its weight $w(u,v,x,y)$ from the weight channel.
The weight represents the confidence associated with the pixel, 
and therefore the relative loss of information due to coding distortion.
Intuitively, distortion of a pixel with higher confidence leads to higher information loss,
and vice versa.

A frame level rate control algorithm should therefore distinguish frames by their relative confidence levels. 
We propose to assign a weight $w_f$ to any perspective frame $f$ with coordinate $(u,v)$, which is 
the average pixel weight across $f$. A unified weight $\tilde{w}_f$ is obtained by rescaling $w_f$ into $[0,1]$ linearly
\begin{align}
w_f = \frac{1}{rh}\sum_{x,y}w(u,v,x,y),~\tilde{w}_f = \frac{w_f}{\displaystyle\max_{f'}w_{f'}},
\end{align}
where $r$ and $h$ are the width and height of $f$, respectively.
The \textit{weighted distortion} $D_f$ of $f$ can thus be defined as a weighted Sum of Square Errors (SSE),
allowing the unified weight to measure the relative information loss
\begin{align}
D^o_f &= \sum_{x,y}(\hat{p}(u,v,x,y) - p(u,v,x,y))^2, \label{eqn:D0_f} \\
D_f &= \tilde{w}^2_fD^o_f, \label{eqn:D_f}
\end{align}
where $D^o_f$ is the ordinary SSE, $\hat{p}$ and $p$ are the decoded and original pixel, respectively. 
The rescaling of $w$ to $\tilde{w}$ makes the order of magnitude of the weight channel immaterial 
and thus allows us to compare different light field images fairly.
In (\ref{eqn:D_f}), $\tilde{w}_f$ is squared to match the squared errors in (\ref{eqn:D0_f}).

\begin{figure}[t]
\begin{center}
\begin{tabular}{cc}
\includegraphics[width=0.5\textwidth]{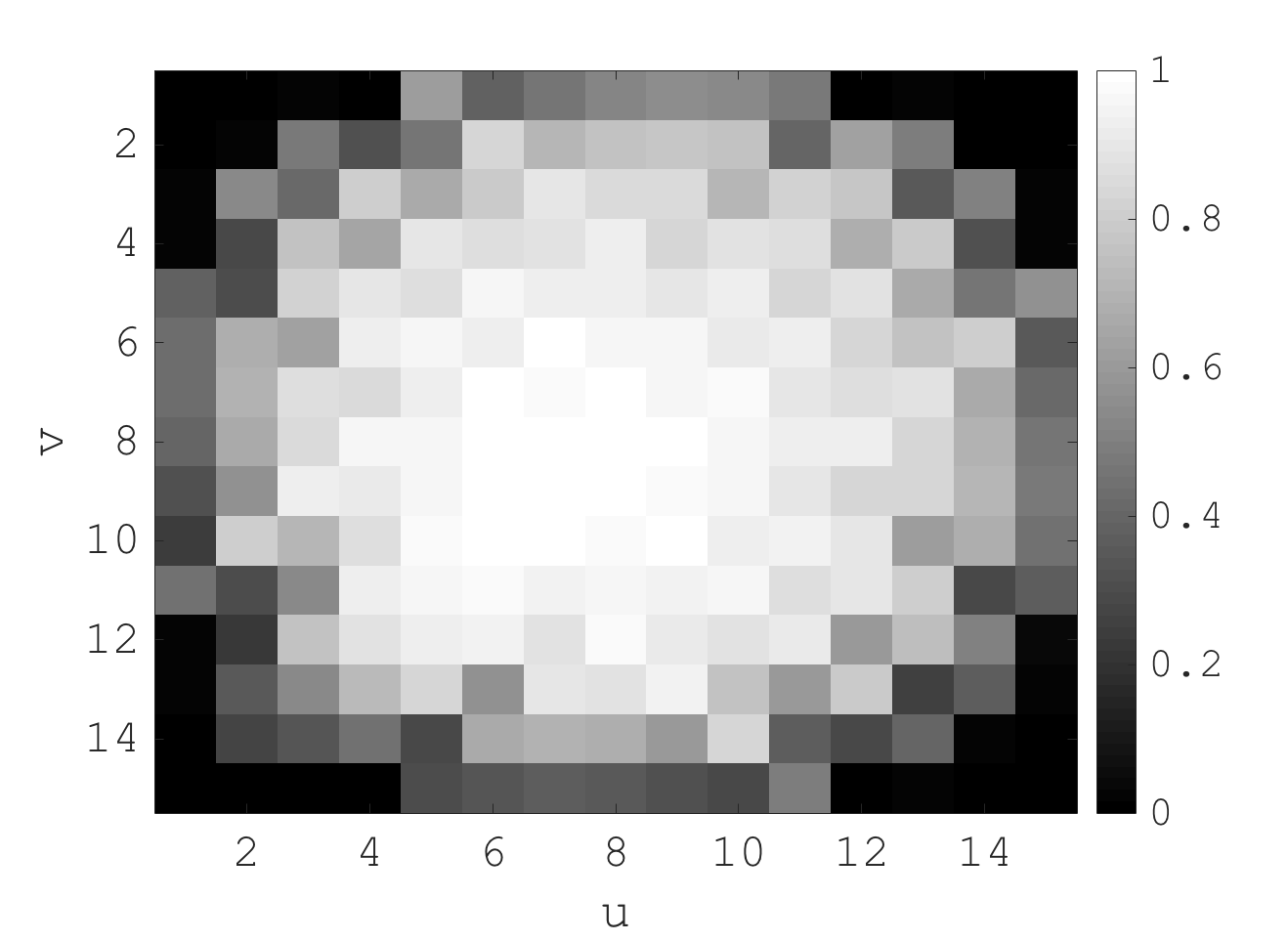} &
\raisebox{.25\height}{\includegraphics[width=0.25\textwidth]{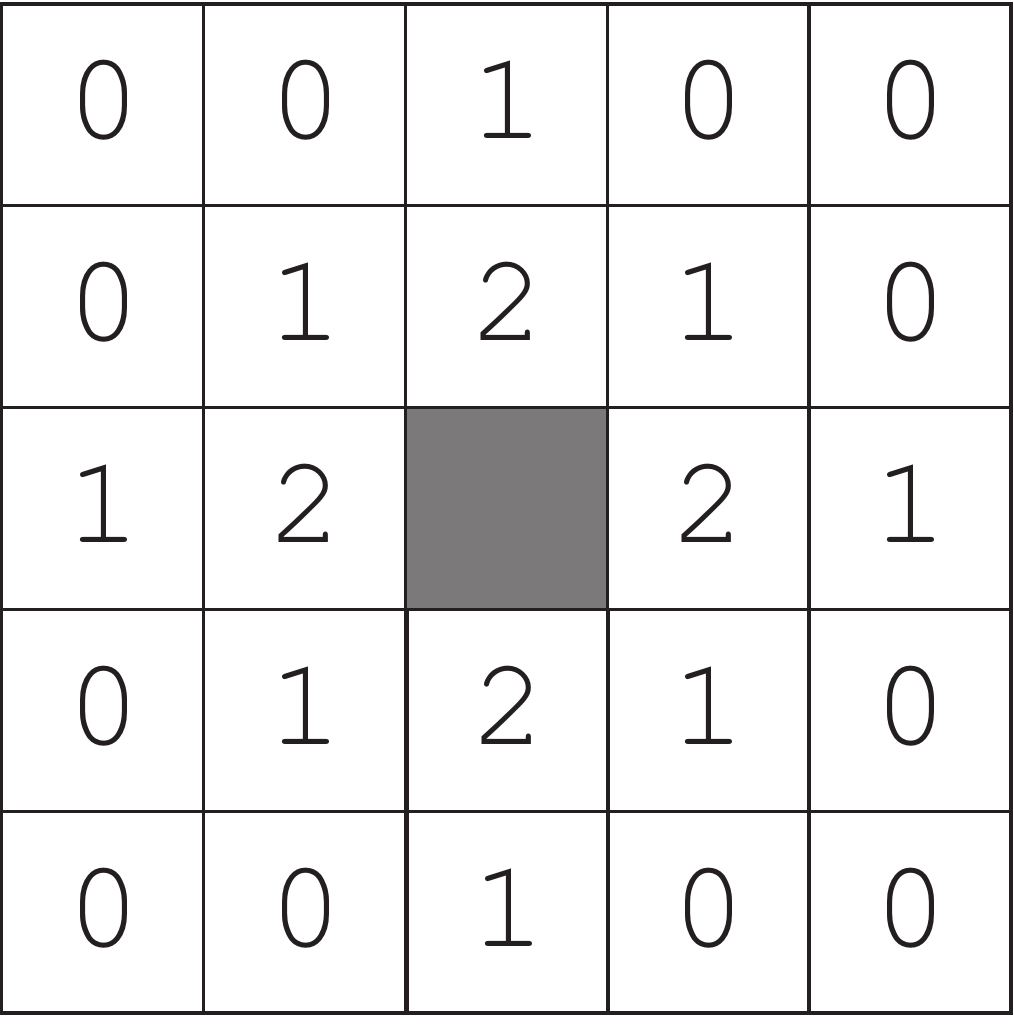}} \\
{\small (a)} & {\small (b)}
\end{tabular}
\end{center}
\caption{\label{fig:weight-delta}%
(a) The weight distribution (b) The proximity function $\delta$}
\end{figure}

A sample distribution of $\tilde{w}_f$ on the $uv$ plane is shown in Fig.~\ref{fig:weight-delta}(a). 
It is a typical pattern with perspective frames in the center of the $uv$ plane have higher weights,
and weights of peripheral perspective frames reduce to zero.

\SubSection{Consistency}
A virtual render of a light field is a synthesized view at an arbitrary position and direction.
The visual consistency of a light field image refers to 
the quality consistency of virtual renders the observer is receiving
when navigating freely in space.
To be precise, since virtual renders of the light field can be obtained by resampling and interpolating  
nearest perspective frames \cite{levoy1996light}, 
the quality of perspective frames needs to be a continuous function of the $(u,v)$ coordinates, 
so that rendered views have smooth transitions when the observer follows a continuous path.

The $uv$ plane with the $\ell 1$ metric, forms a metric space 
where the distance between two perspective frames $f$ at coordinate $(u,v)$ and $f'$ at coordinate $(u',v')$ is
measured as
\begin{align}
d(f, f') = |u-u'| + |v-v'|.
\end{align}
The visual consistency can then be measured by the distortion differences of perspective frame pairs with close distances:
\begin{align} \label{eqn:C}
C = \sum_{f,f'} \delta(f,f') (\min(\tilde{w}_f, \tilde{w}_{f'}) (D^o_f-D^o_{f'}))^2,
\end{align}
where the sum is taken over all perspective frame pairs $(f,f')$.
We shall call $C$ the \textit{discontinuity} term.
Frames with higher confidences shall make larger contribution as they are more ``important",
therefore the unified weights $\tilde{w}$ appear in (\ref{eqn:C}).
The $\delta$ function indicates the proximity of $f$ to $f'$, given by $\delta(f,f')=\max(0, 3-d(f,f'))$. 
Fig.~\ref{fig:weight-delta}(b) shows the possible values of $\delta(f,f')$. $f$ is the gray frame in the center 
and $f'$ can be any white frame, the resulting $\delta$ is marked on $f'$.
It is worth noting that $C$ and $\delta$ can be defined in other forms,
as long as they can be used as a measure for consistency.

\Section{The Bit Allocation Problem}
\SubSection{Problem Formulation}
The goal of our bit allocation framework is to achieve the minimum distortion 
considering the weight and consistency factors, and given the total bit budget.
Using a Lagrange multipler $\lambda$ to control the trade-off between weighted distortion and discontinuity,
one can define the cost function to be minimized as
\begin{align} \label{eqn:T}
\begin{split}
T &= \sum_f D_f + \lambda \sqrt{C} \\
&= \sum_f \tilde{w}_f^2 D^o_f + \lambda\sqrt{\sum_{f,f'}\delta(f,f')(\min(\tilde{w}_f,\tilde{w}_{f'}) (D^o_f-D^o_{f'}))^2}.
\end{split}
\end{align}
The square root is taken over $C$ to improve statistical stability, as norms behave better than 
quadratic forms \cite{cvx} in convex programming.
The total bit budget constraint is simply $\sum R_f \le R$, 
where $R_f$ is the bit cost of frame $f$ and the sum is taken over all perspective frames,
$R$ is the total bit rate budget.

To relate $T$ with $R$ we need to establish the correspondence between $D^o_f$ and $R_f$.
It is clear that $D^o_f$ depends on not only $R_f$, but also the reference frames of $f$.
However, taking into account all dependencies will increase the complexity of the problem tremendously.
In this paper we take the iterative approach as in \cite{Hsu1997}\cite{Sermadevi06}.
$D^o_f$ is assumed to be a function of $R_f$, 
and the resulting solution is used to re-encode the PTS in the next iteration,
until the bit allocation converges.

\begin{figure}[t]
\begin{center}
\begin{tabular}{ccc}
\includegraphics[width=0.3\textwidth]{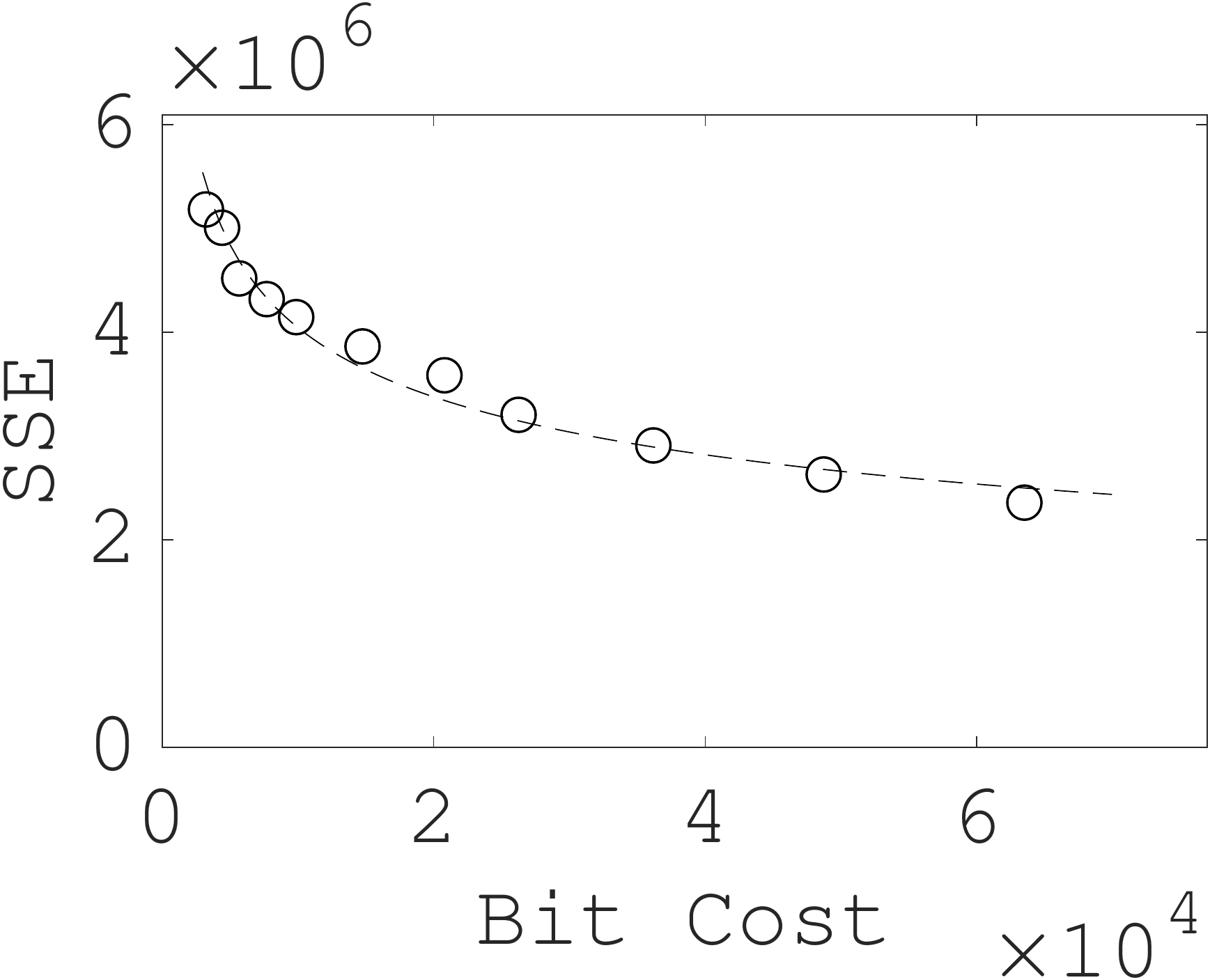} &
\includegraphics[width=0.3\textwidth]{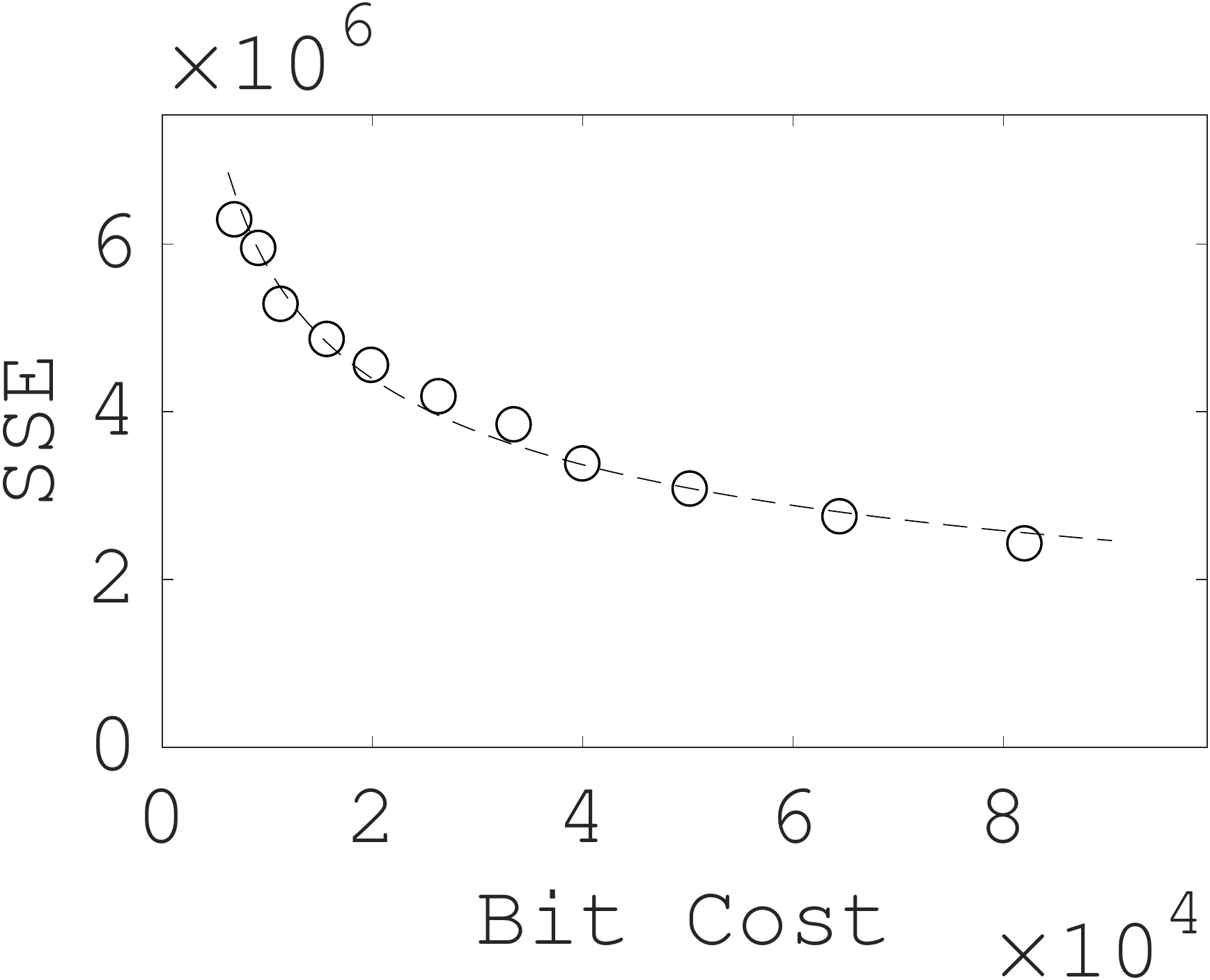} & 
\includegraphics[width=0.3\textwidth]{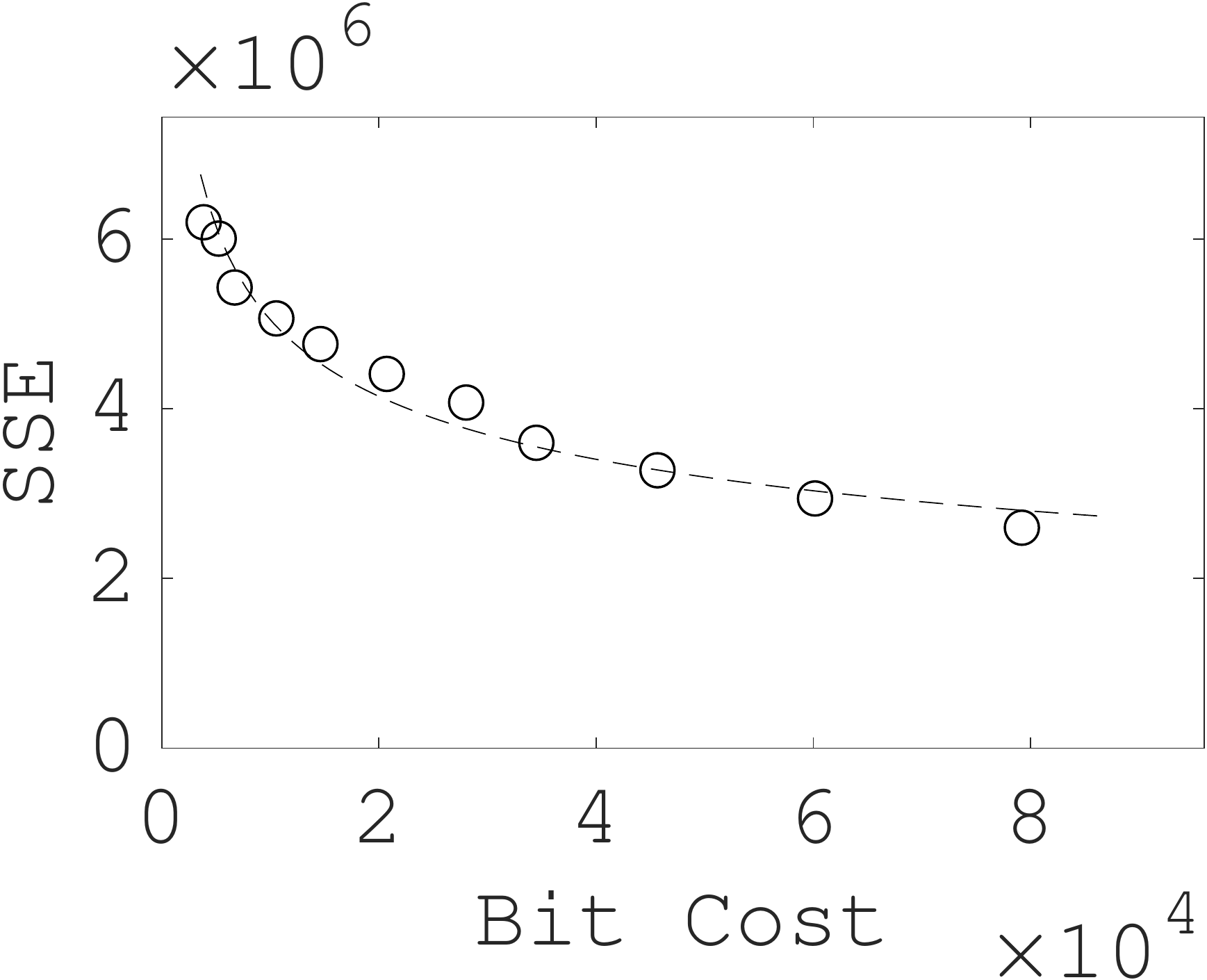}\\
{\small (a)} & {\small (b)} & {\small (c)}
\end{tabular}
\end{center}
\caption{\label{fig:d-r}%
The $D^o_f-R_f$ function}
\end{figure}

The $D^o_f-R_f$ function will be denoted as $D^o_f(R_f)$. 
For three perspective frames \textit{a,b,c} in the light field image \textit{Black\_Fence} \cite{rerabek2016new} , 
their reference frames are fixed and we change the frame QPs
to obtain the $D^o_f-R_f$ relationship shown in Fig.~\ref{fig:d-r} as circles.
The observation that $D^o_f(R_f)$ resembles a monotonic and convex function
enables us to solve the bit allocation problem with convex programming.
Indeed, for each perspective frame $f$ we approximate $D^o_f(R_f)$ by the convex power function
\begin{align} \label{eqn:D-R}
D^o_f(R_f) = \alpha_f R_f^{\beta_f},
\end{align}
where $\alpha_f>0$, $\beta_f<0$. 
The approximated functions for the three perspective frames are drawn as dash lines in Fig.~\ref{fig:d-r}.
Table \ref{tab:linear-fit} shows the parameters $\alpha$, $\beta$ and the $R^2$ coefficients obtained through linear fitting.
It is clear that (\ref{eqn:D-R}) is a good approximation of the $D^o_f(R_f)$ function.

\begin{table}[t]
\begin{center}
\caption{\label{tab:linear-fit}%
$\alpha$, $\beta$ and correlation coefficients obtained through linear fitting}
{
\renewcommand{\baselinestretch}{1}\footnotesize
\begin{tabular}{|c|c|c|c|c|c|}
\hline
Frame &$\alpha$ &$\beta$ &$R^2$ \\
\hline
a &$4.46\times 10^7$ &$-0.261$ &$0.977$  \\
b &$1.96\times 10^8$ &$-0.383$ &$0.985$   \\
c &$6.93\times 10^7$ &$-0.284$ &$0.969$ \\
\hline
\end{tabular}}
\end{center}
\end{table}

Once we have (\ref{eqn:T}) and (\ref{eqn:D-R}), 
our goal to minimize the cost function $T$ can be written as an optimization problem
\begin{align} \label{eqn:opt}
\begin{split}
\text{minimize}~ & \sum_f \tilde{w}_f^2 \alpha_fR_f^{\beta_f} + \lambda\sqrt{\sum_{f,f'}\delta(f,f')(\min(\tilde{w}_f,\tilde{w}_{f'}) (\alpha_fR_f^{\beta_f}-\alpha_{f'}R_{f'}^{\beta_{f'}}))^2}, \\
\text{subject to}~ & \sum_f R_f \le R, R_f \ge 0 \text{ for all } f.
\end{split}
\end{align}

\SubSection{Solving the Optimization Problem}
If we arrange the perspective frames in coding order $(f_1, f_2, ..., f_N)$,
and use $r$ to denote the vector $(R_{f_1}, ..., R_{f_N})$,
(\ref{eqn:opt}) shows that $r$ shall lie inside the $N$-simplex $\lVert r \rVert_1 \le R$, $r\succcurlyeq 0$.
The $N$-simplex is a convex set, 
however the objective to be minimized in (\ref{eqn:opt}) is not convex with respect to $r$.

To make (\ref{eqn:opt}) tractable, we take advantage of the fact that
the discontinuity term in (\ref{eqn:T}) shall not dominate the weighted distortion term,
as we prioritize a small overall distortion over absolute constant quality.
Thus we can divide the task of solving (\ref{eqn:opt}) into two steps:

1. Solve (\ref{eqn:opt}) by omitting the discontinuity term in the objective to obtain an intermediate solution,

2. Approximate the discontinuity term in (\ref{eqn:opt}) by a convex function of $r$ using the intermediate solution.

By omitting the discontinuity term, the objective becomes the sum of weighted 
distortion $T' = \sum \tilde{w}_f^2\alpha_fR_f^{\beta_f}$, which is indeed convex. 
Thus we can solve this optimization problem and obtain an intermediate solution $r_0 = (r_{f_1}, ..., r_{f_N})$ that minimizes $T'$.
Once we have $r_0$, we approximate the discontinuity term $C$ in (\ref{eqn:T}) by 
replacing $D^o_f$ and $D^o_{f'}$ with their first order Taylor approximations:
\begin{align} \label{eqn:D-R-approx}
D^o_f(R_f) \sim \alpha_f(1-\beta_f) r_f^{\beta_f} + \alpha_f\beta_fr_f^{\beta_f - 1}R_f.
\end{align}
Based on our assumption that the weighted distortion should be the dominant term in (\ref{eqn:T}),
the optimal solution of (\ref{eqn:opt}) shall not differ from $r_0$ significantly,
and use (\ref{eqn:D-R-approx}) as a good approximation of (\ref{eqn:D-R}).

Once we use (\ref{eqn:D-R-approx}) instead of (\ref{eqn:D-R}) to compute the discontinuity term,
the optimization problem can be re-written as
\begin{align} \label{eqn:opt-approx}
\begin{split}
\text{minimize}~ & T' + \lambda\lVert Ar+b \rVert_2, \\
\text{subject to}~ & \lVert r \rVert_1 \le R, r \succcurlyeq 0.
\end{split}
\end{align}
where $A$ is a sparse $N^2\times N$ matrix and $b$ is a $N^2\times 1$ vector, given by (\ref{eqn:C})
and (\ref{eqn:D-R-approx}), each of whose rows corresponds to a pair of perspective frames. 
To be precise, row $k$ corresponds to  $(f_i, f_j)$ where $i=\lceil k/N \rceil$, $j=k-N\lfloor (k-1)/N \rfloor$, and 
\begin{align}
A(k, l) &= \begin{cases}
               \sqrt{\delta(f_i, f_j)}\min(\tilde{w}_{f_{i}},\tilde{w}_{f_{j}})\alpha_{f_i}\beta_{f_i}r_{f_i}^{\beta_{f_i} - 1} & \text{if } l=i,\\
               -\sqrt{\delta(f_i, f_j)}\min(\tilde{w}_{f_{i}},\tilde{w}_{f_{j}})\alpha_{f_j}\beta_{f_j}r_{f_j}^{\beta_{f_j} - 1} & \text{if } l=j,\\
               0 & \text{otherwise},
            \end{cases} \\
b(k) &= \sqrt{\delta(f_i, f_j)}\min(\tilde{w}_{f_{i}},\tilde{w}_{f_{j}})(\alpha_{f_i}(1-\beta_{f_i})r_{f_i}^{\beta_{f_i}}-\alpha_{f_j}(1-\beta_{f_j})r_{f_j}^{\beta_{f_j}}).
\end{align}

It is now clear that the objective in (\ref{eqn:opt-approx}) is convex with respect to $r$, 
and a convex programming solver can be used to solve (\ref{eqn:opt-approx}).
In this paper we used the SDPT3 solver implemented in CVX \cite{cvx}, a package for specifying and solving convex programs.
\SubSection{Iterative Encoding}
Based on the solution of (\ref{eqn:opt-approx}), we propose an iterative encoding system 
that minimizes (\ref{eqn:T}) by controlling the bit allocation during PTS encoding.

In the first iteration, we use the default rate control algorithm of HEVC to encode the sequence.
The QPs of all encoded frames are recorded as $(q_1, ..., q_N)$.
For frame $f_i$, as soon as its QP $q_i$ is decided, 
we run a few trial compressions that estimate the bit cost $R_{f_i}$ and distortion $D^o_{f_i}$ 
by using $q'$ as the frame QP, where $q'$ is set to every integer in $[q_i-K,q_i+K]$. 
These $2K+1$ points allow us to estimate the parameters $\alpha_{f_i}$ and $\beta_{f_i}$ in (\ref{eqn:D-R}).
In this paper we set $K=2$.
The trial compressions do not affect the output as the frame QP is set to $q_i$ in the actual encoding.

Starting from the second iteration, 
the frame QPs $(q_1, ..., q_N)$ of the previous iteration are known, 
as well as the parameters $\alpha_f,\beta_f$ for all frames $f$. 
Therefore the solution of (\ref{eqn:opt-approx}) gives the optimal bit allocation as $(r_1, ..., r_N)$. 
For frame $f_i$, the same trial compressions as the first iteration are conducted,
except that the frame QP for the actual encoding is chosen such that its estimated bit cost is close to $r_i$. 
The chosen frame QPs and the estimated parameters $\alpha,\beta$ are passed to the next iteration,
until encoding results converge.
Similar to \cite{Sermadevi06}, it usually takes 3 to 4 iterations to reach convergence in our experiments.

\Section{Experimental Results}
We selected five lightfield images (\textit{Bikes}, \textit{Stone\_Pillars\_Outside}, \textit{Black\_Fence}, \textit{Fountain\_and\_Vincent\_2} and \textit{Friends\_1}) from the EPFL light field image dataset \cite{rerabek2016new} representing different scenarios. 
They were encoded with the default rate control algorithm and our proposed method, both implemented in HM 16.16,
with four different target bitrates (500kbps, 1Mbps, 2Mbps, 4Mbps) under 30 fps.
The ``Low-Delay P" configuration was selected since random access is not necessary for decoding a light field image.
The spiral scan order was chosen due to its simplicity,
though our method is not specific to any arrangement.

\begin{figure}[t]
\begin{center}
\begin{tabular}{ccc}
\includegraphics[width=0.3\textwidth]{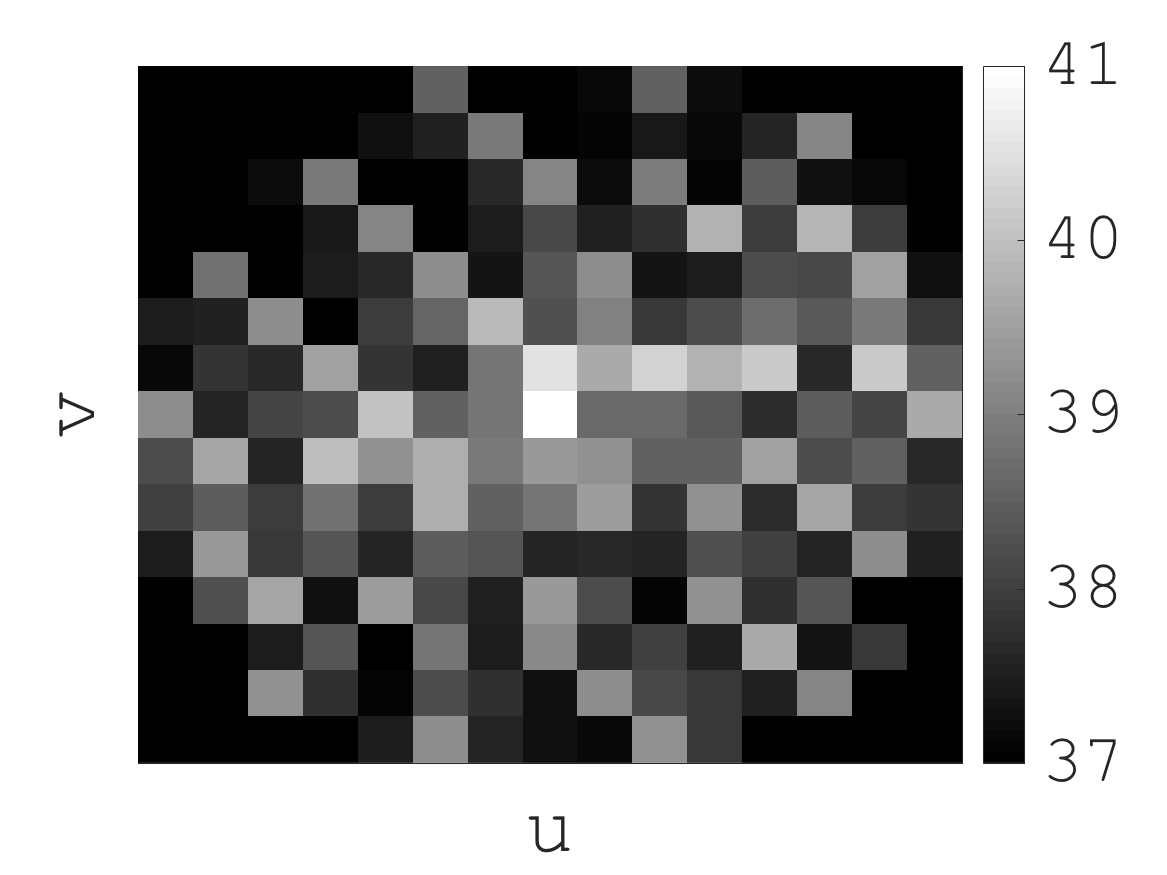} &
\includegraphics[width=0.3\textwidth]{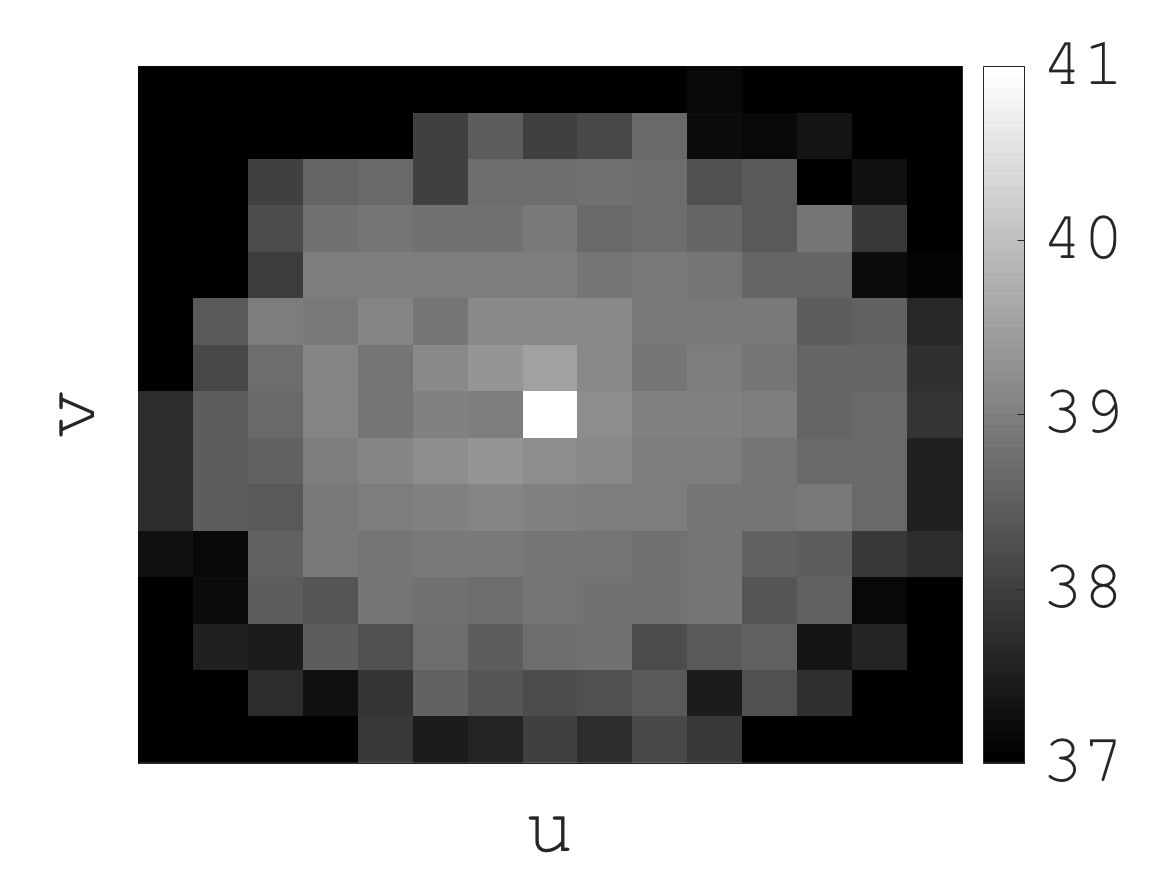} &
\includegraphics[width=0.3\textwidth]{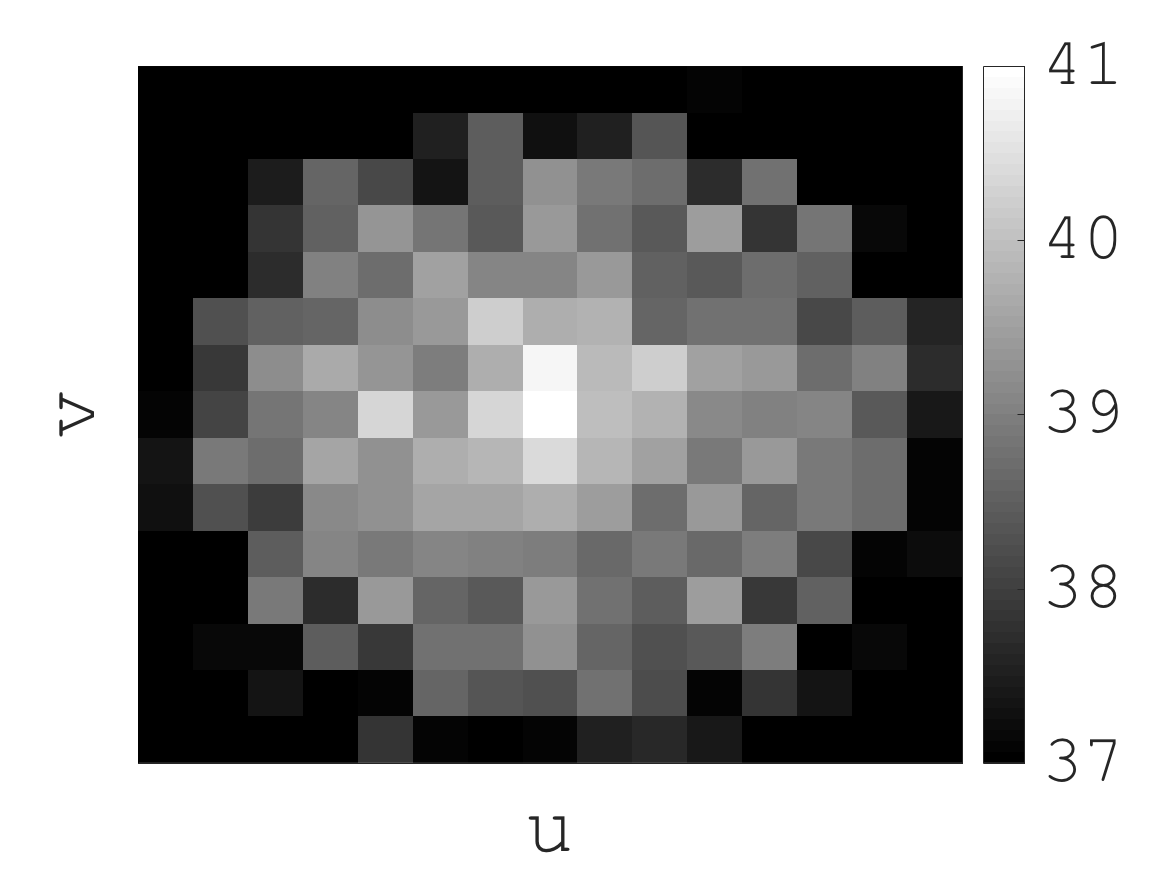} \\
{\small (a)} & {\small (b)} & {\small (c)} \\
\end{tabular}
\end{center}
\caption{\label{fig:psnrs}%
PSNR distributions}
\end{figure}

The Bjontegaard-Delta rate \cite{Bjontegaard} is used to compare the performance of our proposed 
method to the default rate control algorithm. 
However, as our goal is to minimize the cost function (\ref{eqn:T}),
it only makes sense to use (\ref{eqn:T}) to define the MSE in a weighted PSNR metric:
\begin{align} \label{eqn:psnr}
	\text{wPSNR} = 20\log_{10}\bigg(\frac{255}{\sqrt{T/n}}\bigg),
\end{align}
where $T$ is as defined in (\ref{eqn:T}) and $n$ is the number of pixels of the light field image.
Two sets of experiments were carried out where $\lambda$ in (\ref{eqn:T}) was set to $5$ and $0$, respectively. 
The former demonstrates the performance of the overall framework,
and the latter allows us to examine the improvement by considering the weighting factor alone.
The value $\lambda = 5$ was determined heuristically, as it provides a satisfactory trade-off 
between distortion and consistency (see Fig. \ref{fig:psnrs}(b)).

\begin{table}[t]
\begin{center}
\caption{\label{tab:results}%
Experimental results}
\resizebox{\textwidth}{!}{
\renewcommand{\baselinestretch}{1}\footnotesize
    \centering
        \begin{tabular}{|l|c|c|c|c|c|c|c|c|c|c|c|}
        \hline
        \multirow{3}{*}{Image} & & \multicolumn{5}{|c|}{weighting+consistency ($\lambda=5$)} &\multicolumn{5}{|c|}{weighting ($\lambda=0$)}  \\
        \cline{3-12}
        & TBR & \multicolumn{2}{|c|}{default} & \multicolumn{2}{|c|}{proposed} & BD & \multicolumn{2}{|c|}{default} & \multicolumn{2}{|c|}{proposed} & BD \\
        \cline{3-6} \cline{8-11}
         &  & BR & wPSNR & BR & wPSNR & rate & BR & wPSNR & BR & wPSNR & rate \\
        \hline
        \multirow{4}{*}{\textit{Bikes}} 
        & 0.50 &0.51 &36.03 &0.50 &36.58 &\multirow{4}{*}{-16.6\%} & 0.51 &36.92 &0.51 &37.00 &\multirow{4}{*}{-3.3\%}\\
        & 1.00 &1.02 &37.84 &0.99 &38.32 & &1.02 &38.78 &0.99 &38.76 &\\
        & 2.00 &2.02 &39.90 &1.93 &40.29 & &2.02 &40.85 &1.91 &40.79 &\\
        & 4.00 &4.03 &42.35 &3.83 &42.69 & &4.03 &43.25 &3.79 &43.22 &\\
        \hline
        \multirow{4}{*}{\textit{Pillars}} 
        & 0.50 &0.51 &36.30 &0.51 &36.90 &\multirow{4}{*}{-18.5\%} &0.51 &37.26 &0.51 &37.38 &\multirow{4}{*}{-5.8\%}\\
        & 1.00 &1.01 &37.97 &0.98 &38.49 & &1.01 &38.94 &0.98 &38.96 &\\
        & 2.00 &2.01 &39.89 &1.90 &40.30 & &2.01 &40.88 &1.89 &40.90 &\\
        & 4.00 &4.02 &42.12 &3.74 &42.36 & &4.02 &43.05 &3.73 &43.07 &\\
        \hline
         \multirow{4}{*}{\textit{Fence}} 
        & 0.50 &0.51 &37.78 &0.51 &38.76 &\multirow{4}{*}{-24.7\%} &0.51 &38.85 &0.51 &39.19 &\multirow{4}{*}{-11.9\%}\\
        & 1.00 &1.01 &39.62 &0.97 &40.41 & &1.01 &40.70 &0.95 &40.90 &\\
        & 2.00 &2.01 &41.68 &1.91 &42.31 & &2.01 &42.73 &1.90 &42.95 &\\
        & 4.00 &4.01 &43.98 &3.83 &44.47 & &4.01 &45.00 &3.77 &45.17 &\\
        \hline
        \multirow{4}{*}{\textit{Vincent}} 
        & 0.50 &0.51 &35.44 &0.53 &36.05 &\multirow{4}{*}{-24.2\%} &0.51 &36.27 &0.53 &36.51 &\multirow{4}{*}{-5.4\%}\\
        & 1.00 &1.01 &36.50 &1.04 &37.26 & &1.01 &37.62 &1.04 &37.80 &\\
        & 2.00 &2.01 &38.12 &1.96 &38.89 & &2.01 &39.35 &2.00 &39.49 &\\
        & 4.00 &4.00 &40.79 &3.84 &41.34 & &4.00 &41.89 &3.85 &41.96 &\\
        \hline
        \multirow{4}{*}{\textit{Friends}} 
        & 0.50 &0.51 &39.10 &0.50 &39.47 &\multirow{4}{*}{-13.3\%} &0.51 &39.84 &0.50 &39.78 &\multirow{4}{*}{-2.9\%}\\
        & 1.00 &1.01 &40.68 &0.99 &41.02 & &1.01 &41.43 &0.98 &41.41 &\\
        & 2.00 &2.02 &42.33 &1.97 &42.60 & &2.02 &43.08 &2.01 &43.15 &\\
        & 4.00 &4.02 &44.27 &3.99 &44.50 & &4.02 &45.01 &4.02 &45.12 &\\
        \hline
        Average & \multicolumn{5}{|c|}{-} &-19.5\% & \multicolumn{4}{|c|}{-} &-5.9\% \\
        \hline
        \end{tabular}
}
\end{center}
\end{table} 

Table \ref{tab:results} shows the resulting bitrates, weighted PSNRs, and BD-rates obtained from the 
experiments. In the table, ``TBR'' stands for target bitrate, ``BR'' stands for actual bitrate, both are in Mbps.
The weighting factor alone contributes an average 5.9\% and up to 11.9\% BD-rate reduction,
while by jointly considering weighting and consistency,
our proposed method achieves an average 19.5\% and up to 24.7\% BD-rate reduction.

Fig. \ref{fig:psnrs} is an example showing the distribution of perspective frame qualities, using the usual PSNR metric, 
obtained by encoding \textit{Stone\_Pillars\_Outside} with the default algorithm (a) and 
our method where $\lambda=5$ (b) and $\lambda=0$ (c), respectively.
(a) shows that the original HM encoder results in high discontinuity as frames with large PSNR differences are interlaced.
The pattern in (b) shows smooth quality transition, with the exception of the central frame as it is the only I-frame.
The PSNRs fall off from the center to the boundary, in a similar pattern as weights (see Fig. \ref{fig:weight-delta}(a)),
which proves the effectiveness of our joint model.
The consistency factor is disabled in (c),
but the weighting model is still effective and leads to a similar pattern as the weight distribution.
Moreover, as discontinuity is out of concern, (c) results in higher overall discontinuity comparing 
to (b), but is able to achieve higher PSNRs in the central area.

\Section{Conclusions}
In this paper, we proposed a novel bit allocation framework by modeling 
the weighted distortion and consistency for light field images.
The optimal bit allocation is deduced using convex optimization, 
and a weighted PSNR metric is defined for measurement.
Experimental results proves our framework's capability to allocate bits according to the weight distribution, 
and to produce encoded images with high consistency, where a maximum of 24.7\% BD-rate reduction is achieved.

Future work includes improving the rate-distortion model (\ref{eqn:D-R}),
accelerating the iterative process with a fast QP selection algorithm,
as well as a CU level extension of the proposed framework.

\Section{Acknowledgements}
This work was supported by the Natural Science Foundation of China (Project Number 61521002) and Shenzhen Boyan Technology Ltd.

\Section{References}
\bibliographystyle{IEEEtran}
\bibliography{refs}

\end{document}